\documentclass[aps,pre, twocolumn, groupedaddress]{revtex4-1}

\usepackage{graphicx}
\usepackage{dcolumn}
\usepackage{amsmath}    
\usepackage{amssymb}
\usepackage{bm} 
\usepackage{hyperref}
\usepackage{latexsym}
\usepackage{verbatim}
\usepackage{color}
\def\beq{\begin{equation}}
\def\eeq{\end{equation}}

\def\etal{{\it et al.}}

\def\beq{\begin{equation}}                           
\def\eeq{\end{equation}}                           
\def\bea{\begin{eqnarray}}                           
\def\eea{\end{eqnarray}}        

\begin{document}



\title{{Boundary induced convection in a collection of  polar self-propelled particles}}


\author{Shradha Mishra}
\email[]{smishra.phy@itbhu.ac.in}
\affiliation{Department of Physics, Indian Institute of Technology (BHU), Varanasi, India 221005}
\author{Sudipta Pattanayak}
\affiliation{S N Bose National Centre for Basic Sciences, J D Block, Sector III, Salt Lake City, Kolkata 700098}

\date{\today}

\begin{abstract}
{We study a collection of  polar self-propelled particles  
confined to a  long two-dimensional channel. 
We write the  coupled hydrodynamic equations of motion 
for density and polarisation order parameter.
At two confined boundaries, density is fixed to the mean and orientation is anti-parallel with fixed magnitude of polarisation. 
 Such boundary conditions 
 make our system similar to a sheared suspension of self-propelled particles,
 which has many practical  applications.
Antiparallel alignment 
at the two confined boundaries and alignment inside the channel create 
{\it rolls} 
of orientation along the long axis of the channel. For zero self-propulsion speed, density and 
orientation fields are decoupled and density
 remains homogeneous inside the channel.  
For finite self-propelled speed,  density inhomogeneities develop
and these {\it rolls} move 
along the long axis of the channel.  
Density inhomogeneity increases sharply with increasing the self propulsion speed 
and then reaches a 
maximum and again decreases for very large speeds. 
Formation of {\it rolls}  is very  similar to the classic problem
of Rayleigh-Benard convection in fluid dynamics.}
\end{abstract}
\maketitle
\section{Introduction \label{introduction}}
Collective behaviour of active particles are extensively studied 
in  \cite{sriramrev1, sriramrev2, vicsekrev, sriramrev3}.
Large collections of living organisms are known to exhibit highly coherent collective
motion  \cite{vicsek1995, vicsek2, tonertu, traffic, shradhapre}
 This behavior,
often referred to as ``flocking'' spans an enormous range of
length scales and is seen in diverse systems  
 \cite{animalgroup, helbing, feder, kuusela31, hubbard, rauch, benjacob,  harada, nedelec, schaller, vnarayan}. 
These systems are regoursly  studied in {\it bulk} either 
(i) using hydrodynamic equations of motion for slow variables 
(ii) or  microscopic rule based models 
{\it viz.}: Vicsek's model \cite{vicsek1995}.
 {\it But} most 
biological systems are confined to thin geometry \cite{examples}. 
Confinement and boundary 
plays an important role in variety of problems in biology \cite{examples}, 
sheared systems \cite{sheared} 
and other places like in fluid dynamics. One classic example include 
Rayleigh-Benard (RB) convection in fluid \cite{rb}.
In these confined systems, the effect of boundaries are very important. \\
\begin{figure*}[htbp]
\begin{center}
      \includegraphics[height=9cm, width=19cm]{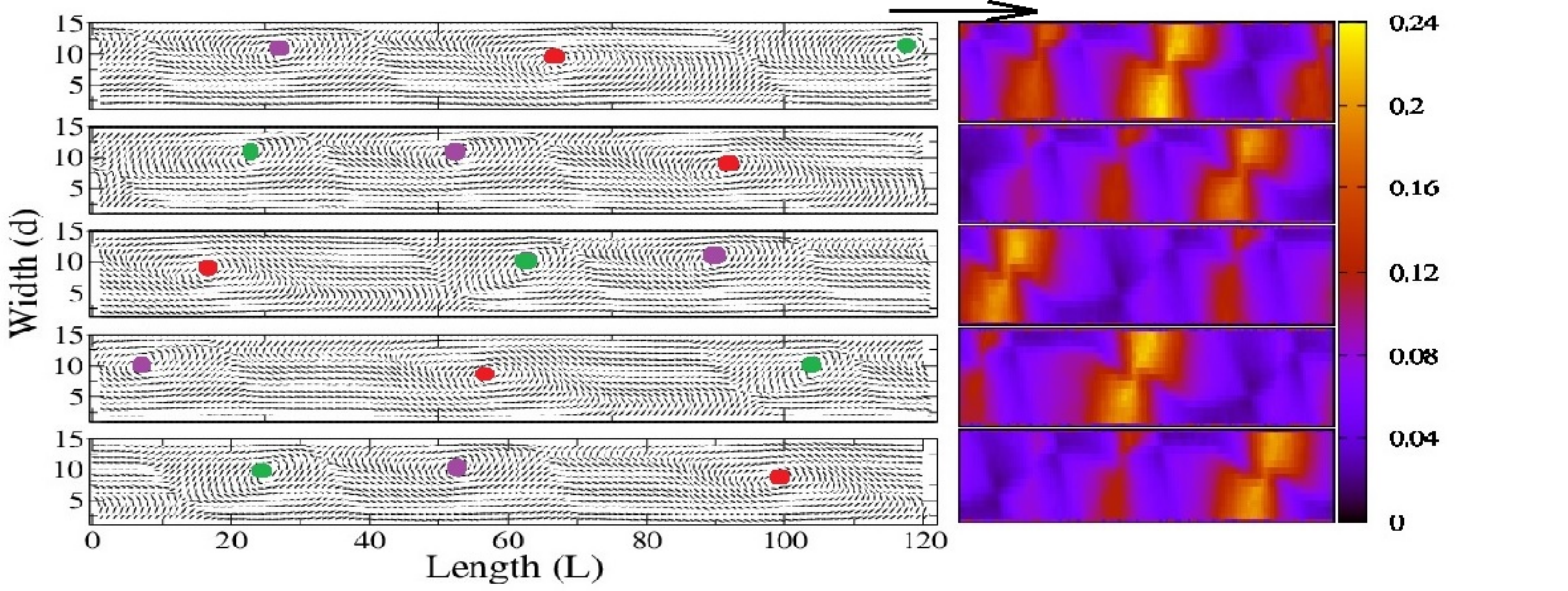}
\caption{(Color online) (left) Vector plot of local polarisation   and (right) density 
inside the channel for  activity $R_A=0.67$.
Different plots are snapshot of polarisation and density at different times. (left) local polarisation
shows vortex type periodic pattern ({\it rolls}) along the long axis of the channel. Different color dots
on periodic {\it rolls} represent distinct vortex. Density also shows periodic pattern.
 Bright regions are high density and dark regions
are low density. Top to bottom figures are from small to large time. With time periodic {\it rolls}  for both density and 
and local polarisation moves from one end to other end of the channel. Arrow on the top of the figure represent direction of
motion of periodic pattern. This direction is spontaneously chosen from two equally possible direction in the system.}
\label{fig1}
\end{center}
\end{figure*}
Boundary can play very important role in a collection of self-propelled particles.
It can induce many
interesting phenomena like, in many cases, boundary can induce spontaneous flow inside the channel 
\cite{faradaydiscussion}.  We write the phenomenological  equations 
of motion  for local density and polarisation order parameter for the collection
of polar self-propelled particles Eqs. \ref{eq1} and \ref{eq2}.  
Self-propelled speed (SPS) of the particle
introduces a non-equilibrium coupling between density and 
polarisation. For zero SPS
both density and polarisation are decoupled. 
We solve these equations  in the confined 
geometry shown in Fig. \ref{fig2}. 
At the two boundaries of the channel orientation of rods are antiparallel,
which produces a  gradient along the confinement direction. 
Diffusion tries to make them parallel.
  Hence the competition between above two
 create  {\it rolls} of orientation along the long-axis of the channel. 
For zero SPS these rolls are static and 
density inside the channel is homogeneous.
For non-zero SPS both density and polarisation are coupled and such coupling
produces moving rolls.\\

\begin{figure}[htbp]
  \begin{center}
      \includegraphics[height=2cm, width=8cm]{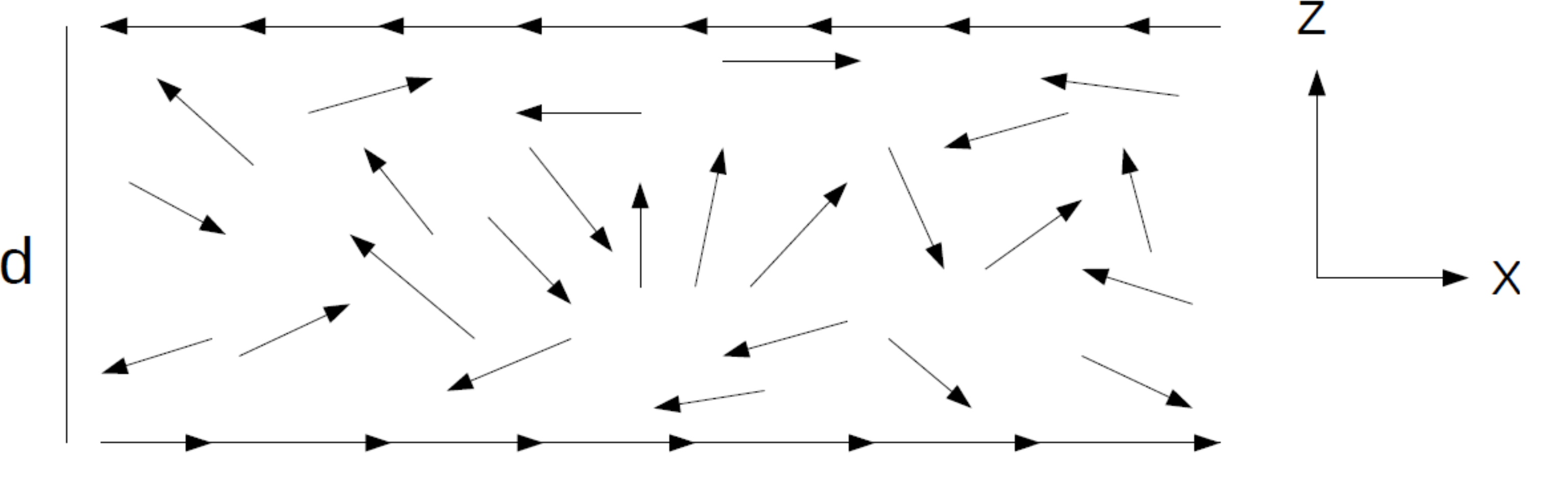}
\caption{Geometry of confined channel and orientation of particle at the two confined boundaries. x-direction
is chosen along the long axis of the channel and z-direction is the confinement direction. Periodic boundary condition is used along the long axis of the channel. Orientation is parallel to +x-direction at bottom boundary (z=1) and parallel to -x-direction at top boundary (z=d). Magnitude of polarisation $|P|=1$ is fixed at two boundaries and density is maintained to  value $\rho_0=0.1$.}
\label{fig2}
\end{center}
\end{figure}

\begin{figure}[htbp]
  \begin{center}
   \includegraphics[scale=0.35]{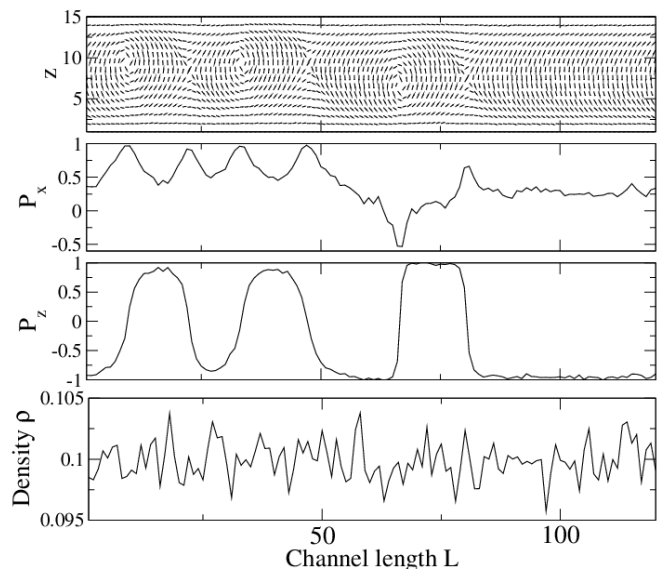}
\caption{For zero activity $R_A=0 $ or SPS $v_0=0.0$:   (a) vector plot of orientation field, which shown periodic vortex type pattern {\em rolls}.  
(b) x-component of polarisation $P_x$ (c) z-component of polarisation $P_z$ and (d) density $\rho(x)$: along the long axis of the channel, averaged over the z-direction.}
\label{fig3}
\end{center}
\end{figure}
In  Fig. \ref{fig1},  we show the  (left) vector plot
of orientation  and (right) 
 density of particles inside the channel for  SPS
 $v_0 = 1.5$ or activity $R_A = 0.67$` at different times. 
We find  inhomogeneous moving  pattern of  orientation and density
 along the long axis of the channel Fig. \ref{fig1}(top to bottom).
Arrow indicate the direction of motion.  \\
In rest of the article,
section 
 \ref{secmodel} discusses the model in detail. Here we also write 
the hydrodynamic equations of motion for density and polarisation. Section \ref{secnumerical}
 discusses the numerical details for solving these equations. We discuss our results
in section \ref{secresult} and finally conclude with discussion and future aspect of 
this study in section \ref{secdiscussion}.
\section{Model \label{secmodel}}
We consider a collection of self-propelled particles of length $l$ confined
to a two-dimensional channel  whose thickness $d$ is  very small compare
to its long axis $L$. 
We fix the length of the channel $L$ and vary the width of the channel $d<<L$.
Orientation at the lower boundary is   parallel to horizontal axis and 
at the upper boundary it is antiparallel and  magnitude of 
polarisation fixed at two boundaries. We also maintain mean density 
at two confined  boundaries to avoid accumulation of particles at 
boundaries.
Periodic boundary condition is used for both density and polarisation 
along the long axis of the channel.  
Geometry of confined channel and orientation of particles at 
the two boundaries is shown in Fig. \ref{fig2}. 

\begin{figure}[t]
  \begin{center}
   \includegraphics[height=7.0cm, width=8.0cm]{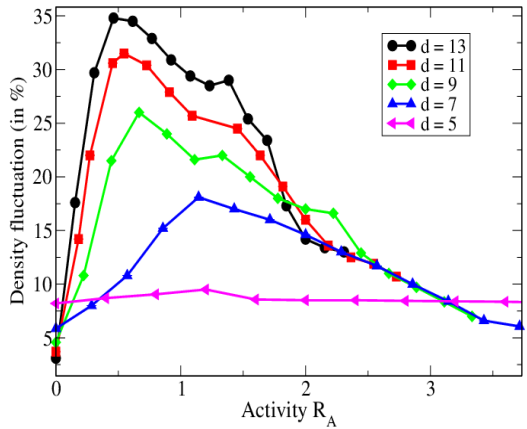}
\caption{(color online)  Plot of percentage  density fluctuation $\%\Delta N$ vs. activity $R_A$ for different 
width $d$ of the channel. $\Delta N$ shows non monotonic behaviour as we
increase activity $R_A$. Density inhomogeneity increases as we increase width of the channel.}
\label{fig4}
\end{center}
\end{figure}
\subsection{Hydrodynamic equations of motion}
Dynamics of the system is described by the  equations of motion for hydrodynamic variables for the 
collection of polar self-propelled particles.
We write the phenomenological coupled hydrodynamic equations
of motion for density $\rho$, because total number of 
particles are conserved and
polarisation $P$, which is an orientation 
order parameter,   is a broken symmetry variable in 
the ordered state.  We write the minimum order terms allowed by symmetry. Two equations are 
\begin{equation}
\frac{\partial \rho}{\partial t} = -v_0 \nabla \cdot (\rho {\bf P}) + {D}_{\rho} \nabla^2 \rho
\label{eq1}
\end{equation}
and
\begin{eqnarray} 
\frac{\partial {\bf P}}{\partial t} &= -D_R( -\alpha_0 + \alpha_1 |P|^2) {\bf P} - \frac{v_0}{2 \rho_0} \nabla \rho + D_P \nabla^2 {\bf P}
\label{eq2}
\end{eqnarray}
Density equation \ref{eq1} is a continuity equation because total number of particles are conserved. Right hand side of Eq. \ref{eq1}
can be written as a divergence of a current ${\bf J}$ which has two parts,
 ${\bf J} = {\bf J}_a + {\bf J}_p$, where
${\bf J}_p \propto \nabla \rho$ is proportional to gradient in density we call it
``passive current'' and
${\bf J}_a \propto v_0 \rho {\bf P}$ is proportional to polarisation vector ${\bf P}$ and
self-propelled speed $v_0$
and we call it as ``active current''. For zero SPS $v_0=0$ or polarisation ${\bf P} =0 $ active
current is zero.\\
 The order parameter equation \ref{eq2} contains  (i) mean field
order disorder terms $\alpha_0$ and $\alpha_1$, (ii) pressure term present
because of density fluctuation and (iii) diffusion in polarisation $D_P$.
 $\alpha_0$ and $\alpha_1$ are 
positive and determine the mean field value of polarisation ${\bf P}$ in the 
bulk
$|P| = \sqrt{\frac{\alpha_0}{\alpha_1}}$. 
We choose $\alpha_0 = \alpha_1 = 1.0$, 
 such choice of $\alpha_0 $ and $\alpha_1$ prefers 
homogeneous polarised steady state  $|{\bf P}| = 1.0$  in the bulk. 
$\nabla \rho$ 
is the pressure term and proportional to the self-propelled 
speed $v_0$ of the particle. 
 $D_P$ term is written in the limit of  equal elastic 
constant approximation for splay and bend deformations in two dimensions and  $D_R$ is the 
rotational diffusion.\\
We  rescale  all lengths by particle length  $l$ (which we choose $1$) and time 
by rotational diffusion time $D_R^{-1}$.
\begin{align}
&r' = r/l; \notag \\
&t' = tD_R.
\label{dimesionless}
\end{align}
and write the dimensionless equations of motion for density 
\begin{equation}
\frac{\partial \rho}{\partial t} = -R_A \nabla \cdot (\rho {\bf P}) + \bar{D}_{\rho} \nabla^2 \rho
\label{eqdensitydimen}
\end{equation}
and polarisation order parameter
\begin{equation}
\frac{\partial {\bf P}}{\partial t} =  (\alpha_0 - \alpha_1 |P|^2) {\bf P} -\frac{R_A}{2 \rho_0} \nabla \rho + \bar{D}_{P}\nabla^2 {\bf P}. 
\label{eqopdimen}
\end{equation}
where 
\begin{equation}
R_A = \frac{v_0}{l D_R} 
\label{eqra}
\end{equation}
 is the dimensionless activity. It is  a ratio between  the self-propelled 
speed $v_0$ and the rotational diffusion $D_R$. Hence we
can increase activity either by increasing  speed $v_0$ or having small $D_R$.
We can also define dimensionless diffusions.
\begin{equation}
\bar {D}_{\rho, P} = \frac{D_{\rho, P}}{l^2 D_R}.
\label{eqdimendiff}
\end{equation}
which is again ratio between bulk diffusion and rotational diffusion.
\section{Numerical Study \label{secnumerical}}
We  numerically solve the coupled  hydrodynamic equations 
of motion for 
density  and polarisation order parameter. 
We go beyond mean field and add Gaussian random white
noise to the density
\begin{equation}
\frac{\partial \rho}{\partial t} = -R_A \nabla \cdot (\rho {\bf P}) + \bar{D}_{\rho} \nabla^2 \rho + \nabla \cdot {\bf f}_{\rho}({\bf r}, t)
\label{eq9}
\end{equation}
and the order parameter
\begin{equation}
\frac{\partial {\bf P}}{\partial t} = - ( -\alpha_0 + \alpha_1 |P|^2) {\bf P} - \frac{R_A}{2 \rho_0} \nabla \rho + D_P\nabla^2 {\bf P} + {\bf f}_{\bf P}({\bf r}, t)
\label{eq10}
\end{equation}
random forces are chosen to have zero mean and correlations
\begin{equation}
<f^i_{\rho}({\bf r}, t) f^j_{\rho}({\bf r'}, t') > = 2 \Delta_{\rho} \delta_{ij}
\delta({\bf r} - {\bf r}') \delta(t-t')
\end{equation}
and
\begin{equation}
<f^i_{{\bf P}}({\bf r}, t) f^j_{{\bf P}}({\bf r'}, t') > = 2 \Delta_{P} \delta_{ij}
\delta({\bf r} - {\bf r}') \delta(t-t')
\end{equation}
where $\Delta_{\rho}$ and $\Delta_P$ are dimensionless  noise strengths.
Numerical study is done for fix noise strength
$\Delta_{\rho} = \Delta_P = 0.05$.
We fix $D_R=0.1$, mean density $\rho_0 = 0.1$, diffusivities $D_{\rho}=D_P=1.0$. Hence activity $R_A$ is 
varied by changing the SPS $v_0$.  Typical diffusive length scale in the 
system $\delta = \sqrt{\frac{D_P}{D_R}} $ and for the above specific  
choice of parameters $\delta = \sqrt{10} \simeq 3.5$. Hence, we choose the lower
limit on $d > \delta$ such that $d$ is not much bigger than $\delta$ and upper limit such that effect
of confinement is important.  Hence $d$ is varied from $5$ to $15$. 
We fixed the length of the channel and SPS is changed from  from zero to large values. \\
We  solve  these PDE's \ref{eq9} and \ref{eq10} using  
Euler method for numerical differentiation
on a grid $\Delta x=1.0$ and $\Delta t =0.1$ (we have checked that numerical
scheme is convergent and stable for the above choice).
Inside the channel we start from initially
random order parameter and  homogeneous  density 
$\rho = \rho_0 =0.1 \pm 0.01$,
At the two boundaries
the magnitude of polarisation is fixed to $P_0 =1.0$ and antiparallel orientation such
that  $P_x(z=1) = 1.0$, $P_z(z=1)=0.0$  and 
$P_x(z=d) = -1.0$, $P_z(z=d)=0.0$ and density is maintained  to mean density 
 $\rho(z=1) = \rho(z=d) = \rho_0$. 
Periodic boundary condition 
is used along the long axis of the channel.\\ 
\begin{figure}[htbp]
  \begin{center}
      \includegraphics[height=9cm, width=9cm]{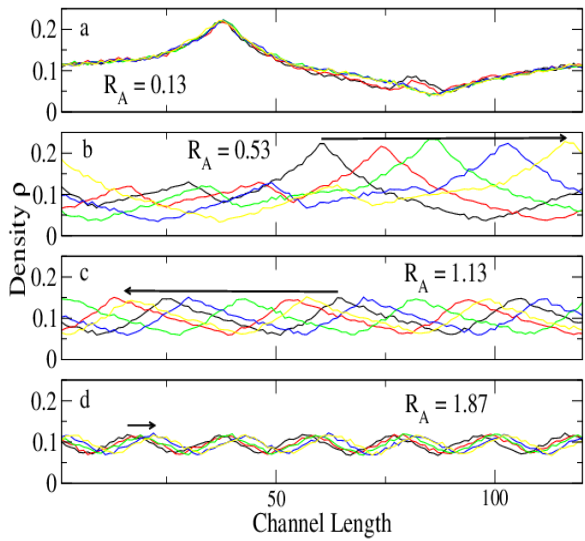}
\caption{(color online) Plot of one dimensional density along the long axis of the
channel for  different self-propelled speeds (a) For small $R_A =0.13 $, 
density shows inhomogeneous structure  but with time this inhomogeneous structure does not move much from
its position. (b)-(c) For $R_A = 0.53 $ and $1.13$,  pattern of high and low density move along 
the long axis of the channel. Patterns move with different speeds at different self-propelled
speeds. (d) Again for large $R_A=1.87$,  pattern does 
not move much with time. Density inhomogeneity
is small compared to that at intermediate $R_A$. Length of the arrow denotes the 
shift in the density pattern with time}
\label{fig5}
\end{center}
\end{figure}
\section{Results \label{secresult}}
We divide our results in two parts. First we discuss what happen when 
activity is turned off, and then we discuss the effect of activity.
\subsection{Zero activity $R_A=0$  or SPS $v_0=0.0$ \label{zerosp}}
For zero activity $R_A=0.0$  density and order 
parameter  are decoupled Eqs. \ref{eq9} and \ref{eq10}. 
Steady state solution for
density is homogeneous density $\rho= \rho_0$.    
Orientation order parameter shows formation of {\it rolls} along the 
long axis of the channel. These rolls are formed because of 
competition between antiparallel boundary condition  which produces a gradient
in orientation and diffusion term  which tries to make them uniform.  Mechanism is
similar to {\em Rayleigh-Benard convection} in fluid dynamics, where 
competition between temperature gradient and gravity produces convective rolls.
In Fig. \ref{fig3} (a) we plot the 
vector plot of orientation field   for
 zero self-propelled speed $v_0=0.0$. 
We find formation of   periodic pattern or {\it rolls} along the 
long axis of the  channel. 
In Fig. \ref{fig3}(b) and (c)  
we also plot the $x$ and $z$ component of orientation field along the long 
axis of the channel which confirms the above periodic structure. In Fig. 
\ref{fig3} (d) we plot the density  along
the long axis of the channel, which remains homogeneous inside the channel.
\subsection{Non-zero activity or SPS $v_0 \ne 0.0$ \label{nonzerospeed}}
When we switch on the activity parameter $R_A$, hydrodynamic equations of motion
for density  and polarisation order parameter Eqs. \ref{eq9} and \ref{eq10} are coupled.
 Non-zero activity introduces an active current ${\bf J}_a \propto v_0 {\bf P} \rho$ 
 which is proportional to the 
SPS $v_0$ and local polarisation ${\bf P}$
(discussed in detail in  section \ref{secmodel}). 
This active current
produces density inhomogeneity and hence enhanced pressure in local
polarisation, which is again proportional to SPS $v_0$ or  activity parameter $R_A$ as shown in Eq. \ref{eq9}
and \ref{eq10}.
\subsubsection{Density fluctuation \label{secdenfluct}}
We first calculate the density inhomogeneity for different 
width of the channel. Density inhomogeneity increases as we increase 
the width of
the channel. In figure \ref{fig4} we plot the percentage density fluctuation $\% \Delta \rho$ 
along the long axis of the channel averaged over transverse direction for five 
different widths $d=5$, $7$, $9$, $11$, $13$ as a function of
activity parameter $R_A$. For very small width $d=5$, density fluctuation is
small ({\em strong confinement}).
For width of the channel $d \ge 7$, density fluctuation shows non-monotonic 
behaviour as a function of activity $R_A$. As we change $R_A$ from zero, first increases
very sharply with a peak at some finite $R_A$ and then decreases slowly for larger activity. 
 Peak position shift towards smaller activity 
as we increase width of the channel. It changes from $R_A = 1.13$ to $0.4$ as we change the 
width from $d=7$ to $d=13$. Hence confinement suppresses the large
density fluctuation present in general in self-propelled
particles in bulk. 
Suppression of density fluctuation because of confinement  is
is also found  previously in the  study of  sheared suspension of Self-propelled
particles  \cite{sheared}.\\
Non-monotonic 
nature of curve gives a finite range of activity, 
where fluctuations are large. For very small activity coupling of density to
background periodic {\em rolls} is small and hence small density
fluctuation. As we increase activity, active contribution to density
current increases: which is proportional to the local order parameter ${\bf P}$ and 
hence background periodic {\em rolls}
of polarisation order parameter.  Activity plays two role here, (i) it produces
an active density current proportional to local polarisation. Since
such active current creates density inhomogeneity hence (ii) it creates pressure in local
polarisation. Hence for very large activity density current will be large but at the
same time pressure will also increase and it will destroy the background 
periodic {\em rolls}. Hence small active current (because ${\bf J}_a \propto {\bf P}$),
hence small density inhomogeneity. \\
\subsubsection{Travelling rolls \label{travelling}}
Even for zero activity as discussed in section \ref{zerosp}, 
antiparallel boundary conditions at the two confined boundaries creates {\em rolls}
 of orientation field. 
For the range of activity when density fluctuation is large, 
these {\it rolls} move from one end to another end of the channel.
We call them as {\it travelling  rolls}: where density and orientation both shows periodic pattern
(please see fig. \ref{fig1} for one such activity $R_A=0.67$). 
In Fig. \ref{fig5} we plot density  for width of the channel $d=13$,
 along the long axis of the channel averaged
over transverse direction. We calculate density for four different times (with equal time difference)
and for   four different activity strength $R_A = 0.13, 0.53, 1.13$ and $1.87$.
For all activities density shows inhomogeneous periodic pattern.  Density inhomogeneity shows variation for
different activity. For very small activity $R_A = 0.13$, density shows small inhomogeneity and 
remains static  with  time. For activity $R_A=0.53$, density is periodic as well as inhomogeneous. These 
periodic pattern move with time. In Fig. \ref{fig5}(b) we draw a horizontal arrow to denote the motion of 
periodic pattern with time. Larger the arrow faster the pattern move. Direction of arrow shows the direction 
of motion of periodic pattern. This direction is spontaneously chosen from two equal 
possible directions in the system. 
For activity $R_A = 1.13$ we get the similar result as for $R_A=0.53$ but density inhomogeneity is weaker.
For activity $R_A=1.87$, density is periodic but inhomogeneity is even more weak and very small shift in peak position 
as a function of time (as shown by small horizontal arrow in Fig. \ref{fig5}(d)). Hence travelling periodic pattern and
density inhomogeneity are coupled and larger density inhomogeneity creates faster moving rolls from one end to
other end of the channel. For  very large activity, density coupling is strong enough such that it destroys the background
periodic pattern of orientation and hence no travelling rolls. \\ 

\begin{figure}[htbp]
  \begin{center}
      \includegraphics[height=4cm, width=8.0cm]{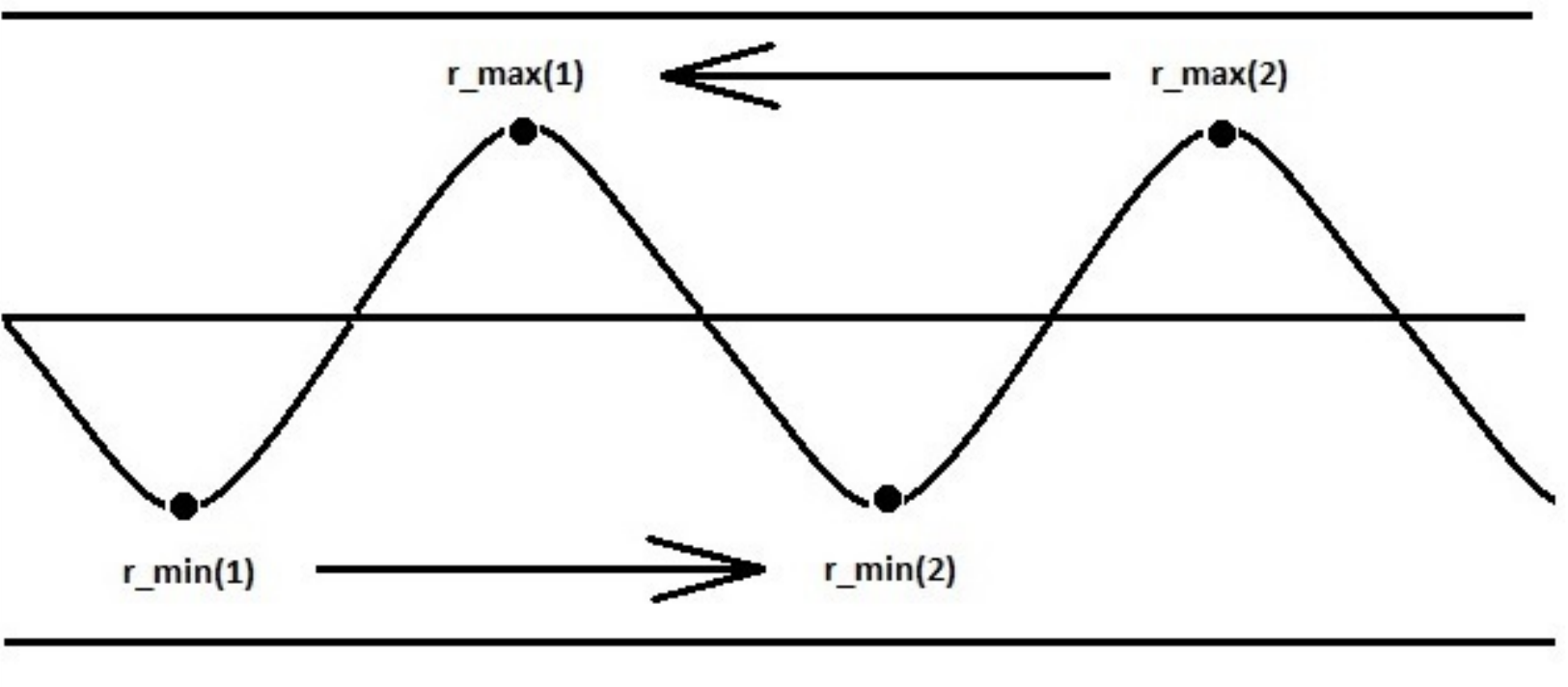}
\caption{Cartoon of periodic pattern of density for finite self-propelled speed $v_0$ or activity $R_A$. $r_{max}[i]$ shows the position of $i^{th}$  maxima and $r_{min}[i]$ the position of the $i^{th}$ minima of density. The two arrows denote the direction of alignment at the two boundaries. We record the position of maxima and minima at different times.}
\label{fig6}
\end{center}
\end{figure}
\subsubsection{Mean square displacement}
\begin{figure}[htbp]
  \begin{center}
      \includegraphics[height=8cm, width=10cm]{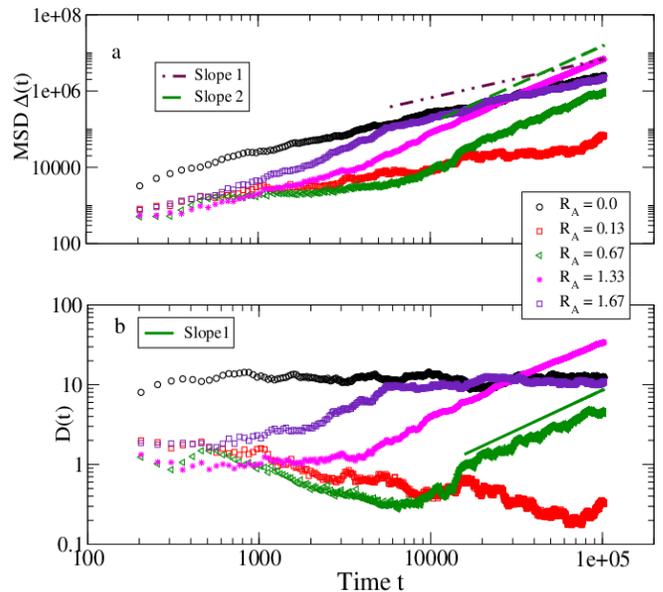}
\caption{(color online) plot of (a) mean square displacement (MSD) $\Delta(t) $  
and (b) corresponding diffusion coefficient $D(t) = \Delta(t)/(2t)$ vs. time $t$,
of position of high and low density peak position
average over many realisation. Different curves are for
different activity ranging from $R_A =0.0$, large values
$R_A=1.67$. For small  $R_A \le 0.13$, $\Delta (t) \simeq t$ (a) and density 
shows diffusive behaviour hence $D(t) $ approaches constant value (b) at large time. 
For intermediate $ 0.67\le R_A <1.33$, $\Delta(t) \simeq t^2$ 
and travelling periodic pattern (a) and hence $D(t) \simeq t$ (b), 
and for large $R_A \ge 1.67$, again 
diffusive and hence $\Delta (t) \simeq t$ (a) and $D(t) $ approaches 
constant value (b). Two Straight lines in (a) are 
line of slope 1 and 2. and in (b) straight line is of slope 1.} 
\label{fig7}
\end{center}
\end{figure}

We further characterise properties of  {\it travelling rolls} assuming  periodic density 
pattern as shown in Fig.  \ref{fig5}.  Cartoon picture of maxima and minima
of periodic density profile for one such realization is shown in Fig. \ref{fig6}. We track the position of maxima and minima
of periodic density pattern.
Position of each maxima and minima we model as position of independent particles moving in one dimension.
Hence each maxima and minima represent one particle and we save the position of 
maxima and minima or particles position with time.
 We calculate the square displacement of such positions and take average 
over all  maximas and minimas and many initial realisations. Hence mean square displacement 
  is defined as
\begin{align}
\Delta(t) & = \frac{1}{N_e}\sum_{n_e=1}^{N_e}\frac{1}{2}\frac{1}{N_i}\sum_{i=1}^{N_i}\big[|r^{n_e}_{max}(i,t_0)-r^{n_e}_{max}(i,t+t_0)|^2 \notag \\
          & +|r^{n_e}_{min}(i,t_0)-r^{n_e}_{min}(i,t+t_0)|^2\big]
\label{msdeq}
\end{align}
where $r^{n_e}_{max/min}(i,t)$ is the  position of maxima/minima of the $i^{th}$ periodic profile 
at time $t$ for $n_e^{th}$ realisation (as shown in Fig. \ref{fig6}). 
Averaging is done over all periodic positions $i = 1, N_i$ and 
number of realisations $n_e =1 , N_e$. For our calculation we used total number of realisation $N_e=20$. \\
When travelling rolls form as discussed in previous section \ref{travelling}, then
$\Delta(t)$ is  proportional to $t^2$. 
In Fig. \ref{fig7}(a) we plot MSD, $\Delta(t)$ 
for different activity $R_A$ for channel of width $d=13$. For zero activity  density is homogeneous 
and no periodic pattern and $\Delta(t) \simeq t$ ({\em diffusive}). 
As we increase activity $R_A=0.13$, MSD
$\Delta(t)$ is  {\it subdiffusion} where $\Delta(t) \simeq t^{\alpha}$ and 
$\alpha <1$. Subdiffusive behaviour shows the arrest of density
in the center of the periodic {\it rolls}, which acts like a  disorder site (with small polarisation).
Similar arrest of density in the presence of quenched disorder field is found
in recent study of \cite{peruaniprl}. But in our model disorder site or center of
the {\it roll} moves for sufficient large activity. For 
activity $R_A \ge 0.67$, MSD  shows  two regimes with initial Subdiffusive with $\alpha<1$ to
later {\it travelling} motion with $\alpha  \simeq 2$.  Hence for $R_A \ge 0.67$,
density remains arrested for some time and then travelling rolls sets in. 
As we further increase activity $R_A=0.67$ and $1.33$
we find initial Subdiffusive with $\alpha<1$ and the  later travelling motion with $\alpha=2$. Time spent
in arrested state decreases as we increase activity.  For
large activity $R_A \ge 1.67$, $\Delta(t) \simeq t^{\alpha}$ for very small time and then switches
to transient faster dynamics and then saturates to diffusive $\Delta(t) \simeq t$ for  large time.\\
In Fig. \ref{fig7}(b) we also plot the diffusivity defined as
\begin{equation}
D(t) = \frac{1}{2t} \Delta(t)
\end{equation}
$D(t)$ remain flat for zero activity, hence {\em diffusion}. For small activity $R_A = 0.13$ it decrease with time
and shows arrested subdiffusion. For intermediate activities  $R_A = 0.67$ to $ 1.33$, $D(t)$
shows initial subdiffusion and later travelling motion with $D(t) \simeq t$. For very large activity $R_A \ge 1.67$
initial  $D(t)$ decreases with time then faster growth to diffusive regime with constant $D(t)$ for very long time.  \\
We further explain the dynamics of particle inside the channel as we vary the activity
$R_A$. For zero activity density and orientation  are decoupled hence we expect normal 
diffusive motion for the particle. As we have discussed before antiparallel boundary 
condition at the two confined boundaries creates periodic ${\it rolls}$ along 
the long axis of the channel. As we switch on activity density is coupled
to periodic orientation field. For very small activity  particles are
trapped in periodic orientation. center of periodic pattern acts as quenched 
disorder site. 
 For intermediate activity
coupling is strong and large density inhomogeneity as shown in Fig. \ref{fig4}. 
Hence for initial transient time particles are trapped to periodic pattern of orientation and 
later active current sets the travelling {\it rolls}. Such initial subdiffusion to propagating 
motion is very common for particle moving in periodic structure. When particle move in a periodic
background it spent some of its time in trapped phase and then start propagation. Similar subdiffusion
to propagation is found for the particle moving in periodic media \cite{shradhajstatphys}.
For very large activity coupling is very strong 
and it destroys the background periodic pattern of orientation and hence diffusive behaviour at late time. 

\section{Discussion \label{secdiscussion}}
In our present work we write the phenomenological  hydrodynamics equations of motion for density and local polarisation
order parameter Eq. \ref{eq9} and \ref{eq10} for the collection of self-propelled particles. We solve these
equation in the confined channel 
of  width $d \ll L$ very small compare to  the long axis of the channel.
 We impose an antiparallel boundary conditions
at the two confinement boundaries and maintain the density at
its mean value and the magnitude of polarisation constant. 
Such a geometry is important because it mimics the shear. 
 First we  solved equations(1,2),  for zero activity.  
Antiparallel boundary conditions impose a gradient of
orientation  along the confinement direction and diffusion  
tries to make them parallel. Hence there is a competition between 
these two terms and   we find periodic 
patterns of the orientation field along the long axis of the channel Fig. \ref{fig3}(a). 
And, since for zero activity density is not coupled to  the orientation field, it remains uniform. \\
Non-zero activity, turn on a contribution of active current, which
is proportional to the local polarisation inside the channel.  Such active currents 
make the inhomogeneous density inside the channel. Density inhomogeneity increases
as we increase the width of the channel. For fixed channel width, as we increase 
activity initially density inhomogeneity increases with activity $R_A$ Fig. \ref{fig4}
and then decrease for large activity $R_A$. 
For fixed channel width, for the range of activity when density inhomogeneity is large, 
 density periodic pattern sets in and start moving from one end 
to other end of the channel. Real space image of moving periodic orientation {\it rolls} 
and density profile for
fixed channel width and activity is shown is Fig. \ref{fig1} for different times.\\
Travelling periodic rolls we observe here is very similar to Rayleigh-Benard convection
in fluids. 
Orientation plays the role of temperature gradient and diffusion  is like gravity which 
acts opposite to gradient. A competition between these two  produces periodic rolls and 
in the presence of activity these rolls move from one
end to other end of the channel.\\
It would be interesting to study other kinds of boundary condition on
the flow properties of active particles inside the channel.
For example recent study of \cite{ncomm}, where boundary induces 
accumulation of particles.
This phenomena where boundary induces spontaneous flow in 
confined channel can give some insight of transport of 
active fluid in biology \cite{biologicaltransport}.\\
\begin{acknowledgments}
S. Mishra would like to thank DST India  for financial support. 
S. Mishra gratefully acknowledge helpful conversation with 
Abhik Basu and Argha Banerjee. 
\end{acknowledgments}



\begin{thebibliography}{10}
\bibitem{sriramrev1} M. Cristina Marchetti {\em et al.}, Rev. Mod. Phys., {\bf 85}, 1143, (2013).
%
\bibitem{sriramrev2} Ramaswamy, S., 2010, Annu. Rev. Condens. Matter Phys. 1, 323
%
\bibitem{vicsekrev} T Vicsek, A Zafeiris, Physics Reports 517 (3), 71-140 (2012).
%
\bibitem{sriramrev3} Toner, J., Y. Tu, and S. Ramaswamy, 2005, Ann. Phys. (Amsterdam) 318, 170
%
\bibitem{vicsek1995} T.\ Vicsek \etal, Phys.\ Rev.\ Lett. {\bf 75}, 1226 (1995).
%
\bibitem{vicsek2} A.Czirok, H.\ E.\ Stanley, and T.\ Vicsek, J.\ Phys.\ A {\bf 30}, 1375 (1997).
%
\bibitem{tonertu} J. Toner and Y. Tu, Phys. Rev. Lett. {\bf 75}, 4326 (1995);
Phys. Rev. E {\bf58}, 4828 (1998),
%
\bibitem{traffic} I. Derenyi, P. Tegzes and T. Vicsek ``Collective transport in locally asymmetric periodic structures'' , Traffic and Granular Flow '97, editors M. Schreckenberg and D. E. Wolf (Springer).
%
\bibitem{shradhapre}Shradha Mishra, Aparna Baskaran, and M. Cristina Marchetti
Phys. Rev. E {\bf 81}, 061916,  (2010). 
%
\bibitem{animalgroup} {\em Three dimensional animal groups}, edited by J. K. Parrish and W. M. Hamner (Cambridge University Press, Cambridge, 1997).
%
\bibitem{helbing} D. Helbing, I. Farkas, and T. Vicsek, Nature (London) {\bf 407}, 487 (2000);
Phys. Rev. Lett. {\bf 84}, 1240 (2000).
%
\bibitem{feder} T. Feder, Phys. Today 60(10), 28 (2007); C. Feare, The Starlings (Oxford University Press, Oxford, 1984).
%
\bibitem{kuusela31} E. Kuusela, J. M. Lahtinen, and T. Ala-Nissila, Phys. Rev. Lett. 90, 094502 (2003).
%

%
\bibitem{hubbard} S. Hubbard, P. Babak, S. Sigurdsson, and K. Magnusson, Ecol. Modell. 174, 359 (2004).
%
\bibitem{rauch}E. Rauch, M. Millonas, and D. Chialvo, Phys. Lett. A 207, 185 (1995).
%
\bibitem{benjacob} E. Ben-Jacob, I. Cohen, O. Shochet, A. Tenenbaum, A. Czirók, and T. Vicsek, Phys. Rev. Lett. 75, 2899 (1995).
%
\bibitem{harada} Y. Harada, A. Nogushi, A. Kishino, and T. Yanagida, Nature (London) {\bf 326}, 805 (1987);
M. Badoual, F. Jülicher, and J. Prost, Proc. Natl. Acad. Sci. U.S.A. {\bf 99}, 6696 (2002).
%
\bibitem{nedelec} F. J. Nédélec, T. Surrey, A. C. Maggs, and S. Leibler, Nature (London) 389, 305 (1997).
%
\bibitem{schaller}
     Schaller V, Weber C A, Semmerich C, Frey E and Bausch A 2010 Nature 467 73; 
    Schaller V, Weber C, Frey E and Bausch A R 2011 Soft Matter 7 3213;
    Schaller V, Weber C A, Hammerich B, Frey E and Bausch A R 2011 Proc. Natl Acad. Sci. USA 108 19183. 
%
\bibitem{vnarayan} V. Narayan \etal, J. Stat. Mech. P01005 (2006);
V. Narayan, S. Ramaswamy, and N. Menon, Science {\bf 317}, 105 (2007).
%
\bibitem{examples}B. Alberts. Molecular biology of the cell, 4th ed., Garland Science, New York, 2002; R. J. Hawkins {\em et al.} , Phys. Rev.
Lett., 102 (2009), No. 5, 058103; 
%
\bibitem{sheared} G. P. Saracco, G. Gonnella, D. Marenduzzo, E.
Orlandini, Phys. Rev. E 84, 031930 (2011); Eur. J. Phys. 10, 1109
(2012).
%
\bibitem{rb} Bodenschatz, E., Pesch, W. and  Ahlers, G., . Annu. Rev. Fluid Mech. {\bf 32}, 709, (2000); Busse, F. H. and Clever, R. M., J. Fluid Mech. {\bf 91}, 319, (1979).
%
\bibitem{faradaydiscussion} A. Zumdieck {\em et al.}, Faraday Discussion, {\bf 139}, 369, (2008); R.Voituriez, J.F. Joanny and J. Prost, {\bf 70}, No. 3, 404, (2005).
%
%
%
%
%
%
%
\bibitem{peruaniprl} O. Chepizhko and F. Peruani
Phys. Rev. Lett. {\bf 111}, 160604, (2013). 
%
\bibitem{shradhajstatphys} S. Mishra,  S. Bhattacharya,
B. Webb and  E. G. D. Cohen, J. Stat. Phys. (2016).
%
\bibitem{ncomm} Antoine Bricard {\it et al.},  	Nature Communications {\bf 6}, 7470 (2015).
%
\bibitem{biologicaltransport} Fournier, R. L. 2007.
Basic Transport Phenomena in Biomedical Engineering,
New York, Taylor and
Francis Group, LLC; Thiriet, M. 2007.
Biology and Mechanics of Blood Flows Part II: Mechanics and Medical Aspects,
Paris, Springer. 
%
\end{thebibliography}
\end{document}